\documentclass[a4paper,twocolumn]{revtex4}
\usepackage{graphicx}
\usepackage{amssymb}
\usepackage{amsmath}
\usepackage{subfigure}
\usepackage{helvet}

\begin{document}

\title{Modeling Worldwide Highway Networks}

\author{Paulino R. Villas Boas} \affiliation{Institute of Physics at
  S\~ao Carlos, University of S\~ao Paulo, PO Box 369, S\~ao Carlos,
  S\~ao Paulo, 13560-970 Brazil}

\author{Francisco A. Rodrigues} \affiliation{Institute of Physics at
  S\~ao Carlos, University of S\~ao Paulo, PO Box 369, S\~ao Carlos,
  S\~ao Paulo, 13560-970 Brazil}

\author{Luciano da Fontoura Costa} \affiliation{Institute of Physics
  at S\~ao Carlos, University of S\~ao Paulo, PO Box 369, S\~ao
  Carlos, S\~ao Paulo, 13560-970 Brazil, National Institute of Science
  and Technology for Complex Systems, Brazil}

\date{\today}

\begin{abstract}
  This letter addresses the problem of modeling the highway systems of
  different countries by using complex networks formalism.  More
  specifically, we compare two traditional geographical models with a
  modified geometrical network model where paths, rather than edges,
  are incorporated at each step between the origin and destination
  nodes.  Optimal configurations of parameters are obtained for each
  model and used in the comparison.  The highway networks of Brazil,
  the US and England are considered and shown to be properly modeled
  by the modified geographical model.  The Brazilian highway network
  yielded small deviations that are potentially accountable by
  specific developing and sociogeographic features of that country.
\end{abstract}

\maketitle

Complex systems are composed of a large number of components obeying
rules that are frequently not well understood. Nevertheless, their
most intrinsic dynamics can be inferred, to some approximation, from
observation of their behavior and used to devise network models
capable of explaining the existing structures and predicting the
network growth and behavior. Special types of complex systems include
human-made structures, such as the Internet, power grids, and highway
networks~\cite{costa2008aam}. These systems are particularly important
because they can provide fundamental clues about human activity and
dynamics and help planning effective and sustainable schemes for
development.  In the current letter, we are interested in the
characterization, classification and modeling of highway networks in
different world regions. Questions of particular relevance which are
addressed in this letter include: (i) can highways be modeled by
single local rules and provide an emergent topology that differs from
random networks?  (ii) are there specific/universal features and
patterns to be found in highways in different countries?  (iii) what
are the optimization processes determining the highway topology?

The terrestrial communication between cities is established according
to an integrated system of railways and highways. These systems give
rise to complex networks optimized to connect nearby cities while
minimizing its overall extension and providing effective
transportation. More specifically, the number of connections of a
single vertex (city) is typically constrained by its proximity to
other vertices, while the establishment of long range connections is
restricted by the distance-dependent cost of edges. Because of these
properties, highways differ from other complex systems by the fact
that they do not present scale-free distribution in the number of
connections and do not exhibit hierarchical structure such as the
World Wide Web~\cite{Ravasz03:PRE}. The several models of geographical
networks that have been developed, e.g.\ the Waxman model for
Internet~\cite{waxman1988rmc}, involve selecting pairs of vertices and
connecting them with probability inversely proportional to the
distance between them. Despite their elegance, such models do not take
into account optimization rules, such as constructing roads that
connect cities found near each connection.

In this letter, a new model of geographical networks recently
introduced in~\cite{Boas09} --- henceforth called the Geographic Path
Network model (GPN) --- is used in modeling of worldwide highway
networks. This model is a generalization of more traditional
geographical networks (e.g.~\cite{hayashi2006rrs}), where cities found
between the extremity vertices have some chance of being incorporated
so that a path, instead of a single edge, is created between the two
reference vertices. The analysis performed in this letter is more
general than that presented in~\cite{Boas09} because highways of three
different countries are investigated and modeled with respect to other
geographical models while taking into account a selection of weighted
measurements. In addition, optimal parameters are obtained for each
model. The three different countries differ with respected to the size
of the highways, the number of cities and, economic development level.
Our analysis shows that the highway networks of Brazil, US and England
can be modeled accurately by an evolving model based on path
transformations~\cite{Boas09}. In fact, the networks obtained by the
application of a relatively simple set of rules are verified to
present topological features substantially similar to those observed
in the real-world networks. Moreover, the rules applied to construct
the networks seem to be universal, no mattering the considered
country, with specific deviations being observed in the case of the
Brazilian network.

\begin{figure*}[!ht]
  \centerline{\includegraphics[width=0.7\textwidth]{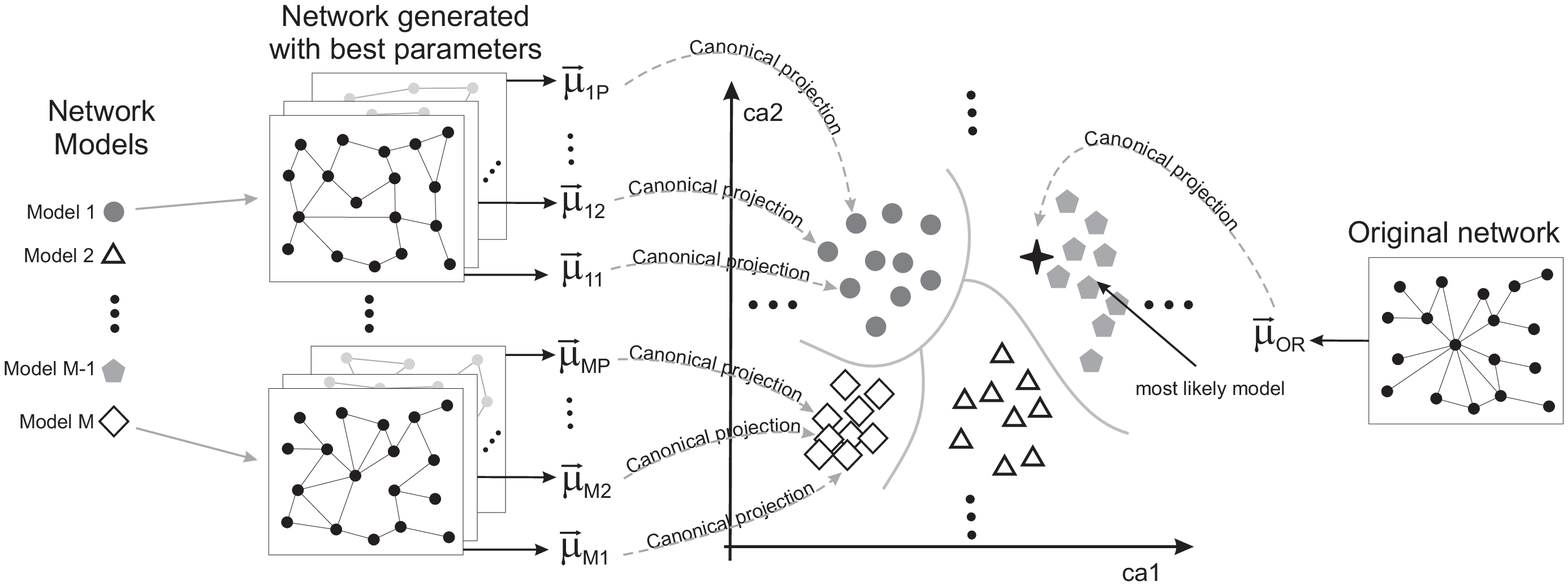}}
  \caption{Methodology for modeling and classification of highway
    networks.  A set of $M$ models is chosen and its best
    configuration of parameters determined, which allows determining
    networks whose topological properties are most similar to the real
    world network. Next, $P$ network realizations with the established
    parameters are generated for each model and a set of measurements
    are estimated and stored into respective feature vectors
    $\vec{\mu}_{ij}$. These feature vectors are then projected into
    two dimensions considering canonical variable analysis (dashed
    gray arrows). Finally, the regions of separation (gray lines) are
    determined by maximum likelihood decision theory. The model
    defining the region in which the real network is projected
    corresponds to the most likely model.}
  \label{fig:methodology}
\end{figure*}

\begin{figure}[!ht]
  \centerline{\includegraphics[width=1\linewidth]{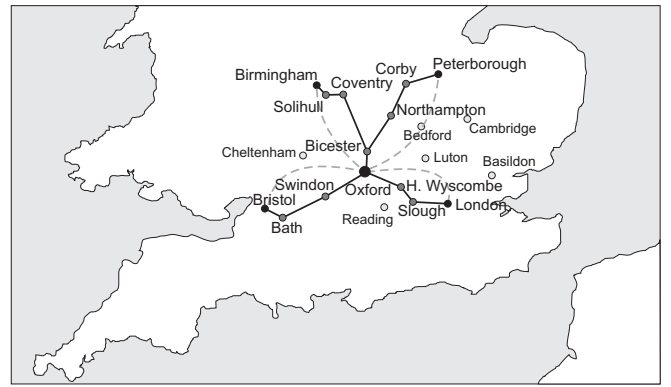}}
  \caption{The model for highway construction. Every city is connected
    to the most populated nearby ones through short paths (solid
    lines) instead of considering direct connections (dashed lines),
    as would be done in traditional geographical complex network
    models. In this case, the starting city, Oxford, is connected to
    London, Bristol, Birmingham, and Peterborough through the paths
    indicated by the solid lines.}
  \label{fig:model}
\end{figure}

Highways are established according to some basic optimization rules,
which are often applied in an empirical fashion. Based on such an
assumption, our model establishes paths that minimize the distance
between cities while trying to maximize the number of nearby cities
that are covered by the incorporated paths.  The model starts with the
isolated cities, and every city is then connected to one or more
destination cities, with priority to the nearest ones, while short
paths should pass through a great number of cities.  The reason why
this rule has been applied is that the destination is generally chosen
by planners to be the most important nearby city and the length of the
chosen path has to be minimized while maximizing the number of covered
cities. By using this procedure, the distance between any pair of
cities and the sum of the length of all highways should be kept
small. It is also reasonably assumed that the importance of a city is
simply given by the size of its population. One example of how cities
can be connected to their destination is presented in
Figure~\ref{fig:model}, with respect to connections from Oxford to
London, Bristol, Birmingham, and Peterborough (dashed lines). The
chosen paths (solid lines) are those with relatively small lengths
which pass through other cities.

More formally, for a origin city $i$, a destination city $j$ is chosen
according to the probability:
\begin{equation}
  P_{ij} \sim \mathrm{Pop}(j)\,e^{-\lambda\,d_{ij}},
  \label{eq:prob}
\end{equation}
where $\mathrm{Pop}(j)$ is the size of the population of city $j$,
$\lambda$ is the only parameter of this model, and $d_{ij}$ is the
geographical distance between $i$ and $j$. If $\lambda$ is large,
there is a higher chance of choosing a destination city nearer the
origin.  For small values of $\lambda$, the chance of choosing a
distant city is increased.  After choosing a destination, a similar
rule for including cities along the path is applied, but those cities
whose distances to either origin or destination are greater than the
distance between the origin and destination are not considered. Then,
the set $S$ of all possible cities including $j$ is sorted according
to the distance from $i$. Starting from $i$, the next city $k$ is
chosen with probability given by Eq.~\ref{eq:prob}, with $j=k$.  After
choosing $k$, all cities with distance from $i$ less than the distance
from $i$ to $k$ are removed from the set $S$, and the next city is
chosen. This procedure is repeated until there is just one element in
the set $S$, which is $j$. The resulting sequence of cities defines
the path from $i$ to $j$.

Overall, the model construction includes two stages. The first one
corresponds to finding a path to a destination for every city. The
second stage is to randomly choose cities and the corresponding path
to another destination until the desired average number of connections
per city is obtained. After completing these stages, the respective
weighted undirected network is obtained, where every city is a vertex,
and two cities are connected through an edge weighted by their
distance if they are neighbors along one of the obtained paths.

In order to characterize the geometrical networks considered in this
article, the following measurements have been used~\cite{Costa07:AP}:
(i) average strength, i.e.\ average distance between neighbor
connected cities; (ii) average of the average strength between the
neighbors of the cities; (iii) Pearson correlation coefficient between
the vertex strength at both ends of the edges; (iv) weighted
clustering coefficient calculated by considering the inverse of the
weight; (v) average shortest path length; (vi) average betweenness
centrality; (vii) central point dominance; (viii) average concentric
degree of level 2 (degree of a vertex is the number of its
connections); (ix) average concentric clustering coefficient of level
2; and (x) average concentric divergence ratio of level 2.  All these
measurements are described and discussed in~\cite{Costa07:AP}, and
only the last three have not been calculated considering the edge
weights.

The GPN model has been evaluated with respect to the highways of
Brazil, US and England.  The respective networks, as well as the
coordinates and the population of the cities, have been compiled
manually from several sources on the Internet. Only federal interstate
highways and the main cities of each country have been considered,
except in the case of England due to its small territory, in which
case all highways have been considered.  The biggest network is the
Brazilian highway network with 487 cities, followed by US with 244
cities, and by England with 136 cities.

The number of inhabitants of each city has also been determined in a
similar way.  Alternative models have been compared to our suggested
GPN model. Although there are many geographical network
models~\cite{boccaletti2006cns}, only those which allow the
specification of the position of the vertices were used in our
analysis. This constraint is necessary since all models used have the
same number of vertices with the same positions as the original
network. The process of building the models is the same: the first
step is to start with a set of disconnected vertices whose positions
are given by the original network. Then, the cities are connected
through rules which depend on each model.  In order to determine the
accuracy of the GPN model, it has been compared to two other
geographical models. The first one is a version of the Waxman
geographical model~\cite{waxman1988rmc}, represented by WGN, in which
the probability of connecting two vertices $i$ and $j$ is proportional
to $\alpha\,e^{-\lambda\,D_{ij}}$, where $\alpha$ is one parameter to
obtain the desired average vertex degree, $D_{ij}$ is the geographical
distance between vertices $i$ and $j$, and $\lambda$ is a parameter of
the model that controls the chance of choosing $j$ near or far away. A
small value for $\lambda$ means that the chance of choosing a vertex
$j$ far away from $i$ is not so small as compared to vertices near
$i$. For higher values, the chance of choosing a vertex far away from
$i$ is reduced.

It has been previously observed~\cite{Boas09:IJBC} that the WGN model
generate networks whose topological features are close to the US
highway network. Indeed, the currently proposed GPN model is a
generalization of the WGN model.  The second model is a geographical
and scale-free (e.g.~\cite{hayashi2006rrs}), represented by GSF, whose
construction is similar to the Barab\'asi and Albert scale-free
model~\cite{Barabasi99:Science}, in which vertices with a small number
of connections are added sequentially, and the probability of choosing
a vertex is proportional to its degree (the number of connections of a
vertex). In the GSF model, however, we start with a set of
disconnected vertices and connect two vertices $i$ and $j$ with
probability proportional to
$e^{-\lambda\,D_{ij}}(k_j+\delta)/\sum_l{(k_l+\delta)}$, where
$\lambda$ and $D_{ij}$ are the same as for WGN, $k_j$ is the degree of
$j$, and $\delta$ is the second parameter of the model which provides
a small chance of choosing vertices that are still isolated.

In order to achieve a fair comparison between the models, the $n$
parameters $\vec{p}=\{p_1,p_2,...,p_n\}$ of each of them have been
adjusted so that they are optimized with respect to each of the three
highway networks. This is obtained by varying all parameters linearly,
through successive approximations, in order to find those which imply
the smallest Euclidean distance~\cite{Costa:book}, which can be
obtained by:
\begin{equation}
  D_{\vec{p}} = \sqrt{\sum_{l=1}^{10}{(\mu_l^{\mathrm{Orig}}-\overline{\mu}_{\vec{p}})^2}},
\end{equation}
where the sum is performed considering the set of 10 measurements
described before, $D_{\vec{p}}$ is the Euclidean distance from the
real highway network to the corresponding model with parameters
$\vec{p}$; $\mu_l^{\mathrm{Orig}}$ is the measurement $l$ for the real
highway network; and $\overline{\mu}_{\vec{p}}$ is the average
measurement $l$ over all realizations of the corresponding model with
parameters $\vec{p}$.

Table~\ref{tab:parameters} presents the best parameters found after
500 realizations of each model for each set of parameters $\vec{p}$.

\begin{table}[!ht]
  \caption{Best parameters found for the models.}
  \label{tab:parameters}
  \vspace{-0.1cm}
  \begin{center}
    \begin{tabular*}{0.33\textwidth}{@{\extracolsep{\fill}}l l r r r}
      \hline
      Model  & Parameters & England & US   & Brazil \\
      \hline
      GWN    & $\alpha$	  & 1.0	    & 1.0  & 0.95   \\
             & $\lambda$  & 16.0    & 17.0 & 27.0   \\
      \hline
      GSF    & $\delta$	  & 1.0     & 1.0  & 1.0   \\
             & $\lambda$  & 41.0    & 36.0 & 39.0   \\
      \hline
      GPN     & $\lambda$  & 32.0    & 19.0 & 44.0   \\
      \hline
    \end{tabular*}
  \end{center}
\end{table}

\begin{figure*}[!ht]
  \subfigure[]{\includegraphics[width=0.3\textwidth]{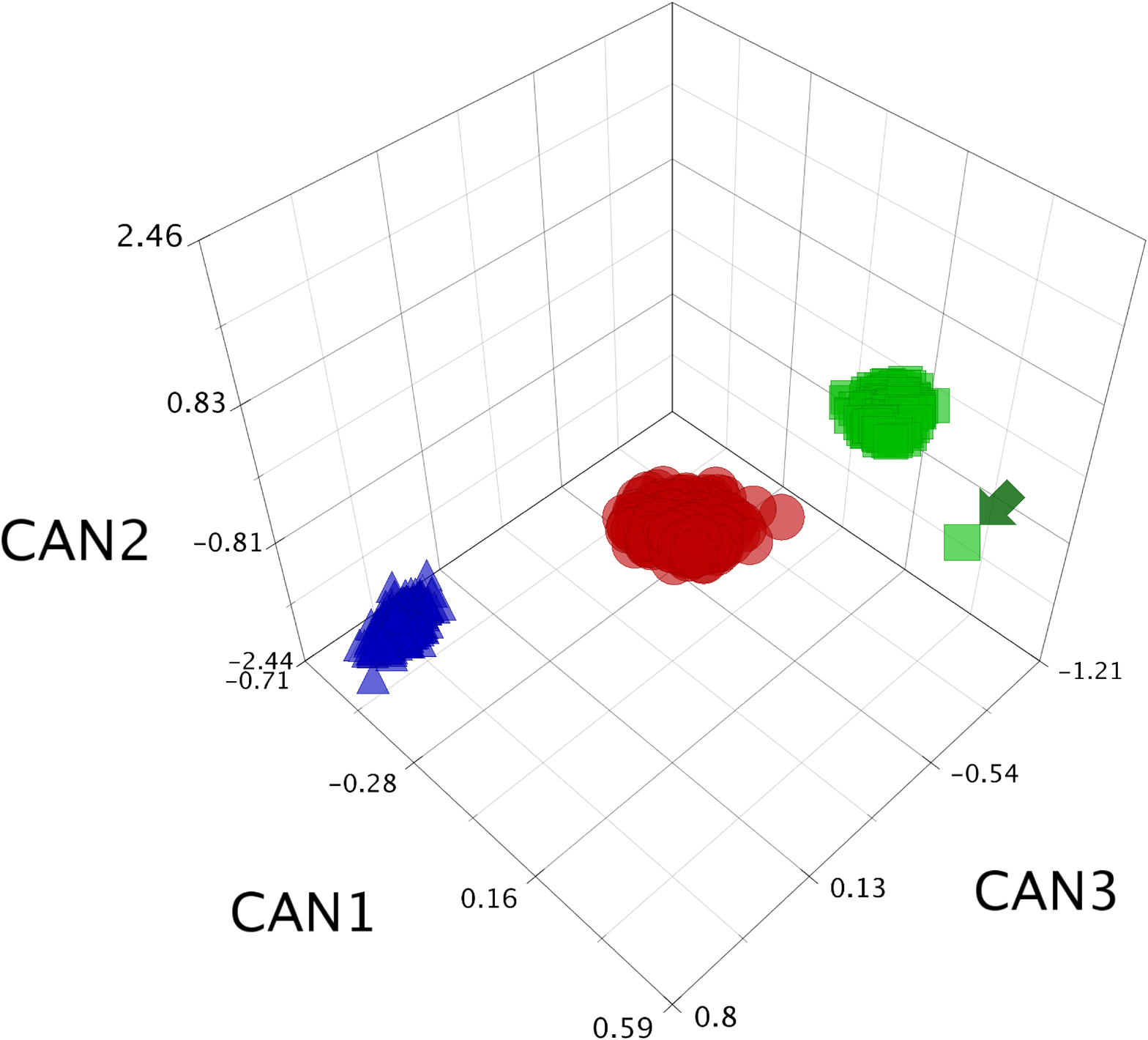}}
  \hspace{0.01\textwidth}
  \subfigure[]{\includegraphics[width=0.3\textwidth]{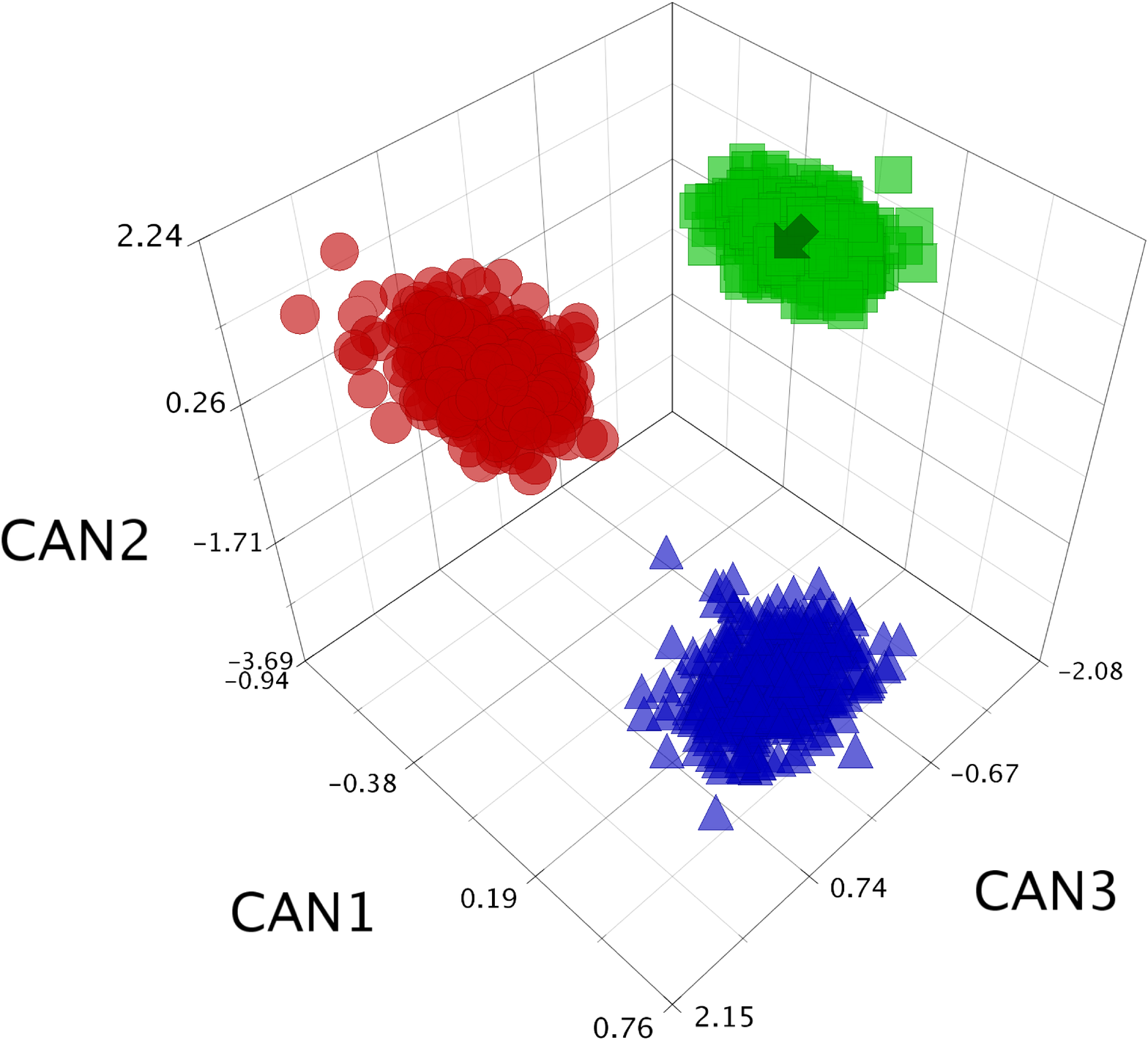}}
  \hspace{0.01\textwidth}
  \subfigure[]{\includegraphics[width=0.3\textwidth]{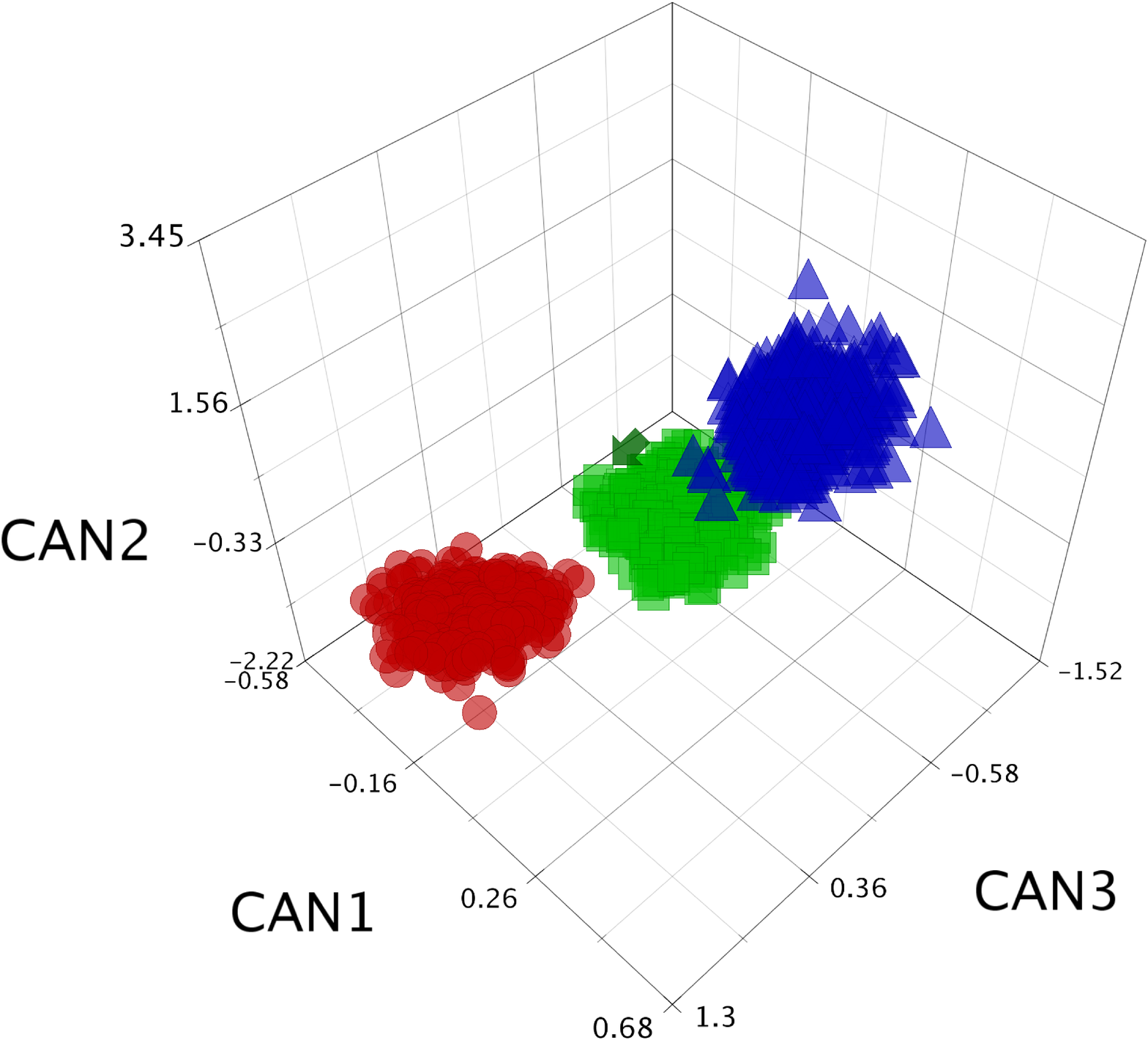}}
  \caption{The 3D phase space defined by the first three canonical
    variable projections for the (a) Brazilian, (b) England, and (c)
    US highway networks (indicated by arrows) and for the models: GPN
    (green square), WGN (red circle), and GSF (blue triangle). The
    highway networks are already assigned with the symbol of the
    respective class obtained by the maximum likelihood methodology.}
  \label{fig:classification}
\end{figure*}

A new set of 500 realizations and the corresponding measurements have
been obtained for each model using the best parameters of
Table~\ref{tab:parameters}. Since these measurements are often
correlated, canonical variable analysis~\cite{Campbell1981gcv,
Duda2000pc} has been applied in order to reduce the correlation
between them, provide a means for respective visualization, and ensure
optimal separation between all the involved categories. The results of
this methodology are shown in Figure~\ref{fig:classification}, where
the classification of the three highway networks is also provided.

The classification is performed by extracting a set of measurements
for each network model realization. These measurements were first
standardized~\cite{Costa:book} (i.e. subtract the mean and divide by
the standard deviation of each class) in order to have zero means and
unit standard deviation. These features were then projected into the
two dimensional space by canonical variable
analysis~\cite{Campbell1981gcv}.  Finally, the classification was
performed by maximum likelihood decision theory~\cite{Costa:book}.
When all categories involve the same number of individuals (as is the
case in this work), the maximum likelihood methodology determines the
probability of each model with respect to a give set of attributes and
associates to the original highway network the model that yields the
maximum likelihood.

The results shown in Figure~\ref{fig:classification} clearly indicate
that the GPN allows the best reproduction of the respective highway
networks for all three countries. The Brazilian highway network
resulted, however, a little bit more distant from the GPN
networks. This result can be a consequence of the specific way in
which the Brazilian highways were constructed and/or the diverse local
geography which includes large forests, uninhabited regions and
swamps.  It is interesting to note that all the parameters involved in
the GPN model are higher for Brazil and smaller for the US
highways. Such a trend is possibly related to the homogeneity of
highway distributions, country development and population
uniformity. For instance, the US network presents a more integrated
highway system, which connect all locations even along deserts.

All in all, our results shown that different worldwide highway system
can be accurately modeled by the simple, possibly universal, rules
embedded in the GPN model.  Slight deviations obtained for the case of
the Brazil network suggests distinguishing topological features which
are potentially related to the developing stage and sociogeographic
specific features.  Additional studies could investigate how the
proposed model reproduces the time evolution of highway networks in
different countries.

\acknowledgments{Luciano da F. Costa is grateful to FAPESP
  (05/00587-5), CNPq (301303/06-1 and 573583/2008-0) for financial
  support. Francisco A. Rodrigues acknowledges FAPESP sponsorship
  (07/50633-9), Paulino R. Villas Boas acknowledges FAPESP sponsorship
  (08/53721-9).}

\bibliographystyle{unsrt}
\bibliography{ref}
\end{document}